\documentclass[fleqn,12pt,twoside]{article}
\usepackage{espcrc1}

\usepackage{graphicx}
\usepackage[figuresright]{rotating}

\newcommand{\gaz}{g_A^{\mbox{$\scriptscriptstyle (Z)$}}}
\newcommand{\run}[1]{\widetilde{\alpha}_{#1}}
\newcommand{\hsp}[1]{\hspace*{#1 mm}}
\newcommand{\smallfrac}[2]{\mbox{\small ${\displaystyle \frac{#1}{#2}}$}}
\newcommand{\footfrac}[2]%

\newcommand{\AmS}{{\protect\the\textfont2
  A\kern-.1667em\lower.5ex\hbox{M}\kern-.125emS}}

\title{Neutrino proton elastic scattering and the spin structure of the
proton}

\author{Steven D. Bass\address[MCSD]{Institute for Theoretical Physics,
University of Innsbruck, 
Technikerstrasse 25, A6020 Innsbruck, Austria}%
        \thanks{Work supported in part by the Austrian Science Fund, 
FWF grant M770.}
}     
\begin{document}

\maketitle

\begin{abstract}
Neutrino proton elastic scattering and polarized deep inelastic
spin sum rules provide complementary information about the spin
structure of the proton.
We outline the two approaches and what they may teach us about
the transition from current to constituent quarks in QCD.
\end{abstract}

\section{Introduction}

Understanding the spin structure of the proton is one of the most 
challenging problems facing subatomic physics:
How is the spin of the proton built up out from the intrinsic spin
and orbital angular momentum of its quark and gluonic constituents ?
What happens to spin in the transition between current and constituent
quarks in low-energy QCD ?
Key issues include the role of polarized glue and gluon topology in 
building up the spin of the proton.

Measurements of the proton's $g_1$ spin structure function in polarized 
deep inelastic scattering have been interpreted to imply a small value 
for the flavour-singlet axial-charge:
\begin{equation}
g_A^{(0)}\bigr|_{\rm pDIS} = 0.15 - 0.35 .
\label{eqa1}
\end{equation}
This result is particularly interesting \cite{spinrev} 
because $g_A^{(0)}$ is interpreted in the parton model 
as the 
fraction of the proton's spin which is
carried
by the intrinsic spin of its quark and antiquark constituents.
The value (\ref{eqa1}) is about half the prediction of
relativistic constituent quark models ($\sim 60\%$).
It corresponds to a negative strange-quark polarization
\begin{equation}
\Delta s = -0.10 \pm 0.04 
\label{eqa2}
\end{equation}
(polarized in the opposite direction to the spin of the proton).
The small value of $g_A^{(0)}|_{\rm pDIS}$ 
extracted from 
polarized deep inelastic scattering has inspired vast 
experimental and theoretical activity to understand the spin 
structure of the proton (and related connections to the axial U(1) problem).
New experiments 
are underway or being planned to map out the proton's 
spin-flavour structure and to measure the amount of spin carried 
by polarized gluons in the polarized proton -- for a review see \cite{bassdr}.

In this paper we explain the spin sum rule connecting $g_1$ and 
$g_A^{(0)}$
and how additional valuable information could be obtained from
a precision measurement of neutrino proton elastic scattering.

\section{$g_1$ spin sum rules in polarized deep inelastic scattering}

Sum rules for polarized deep inelastic scattering are derived starting 
from the dispersion relation for photon nucleon scattering:
\begin{equation}
A_1 (Q^2, \nu)
=
\beta_1 (Q^2) 
+
{2 \over \pi} \int_{Q^2/2M}^{\infty} \ {\nu' d \nu' \over \nu'^2 - \nu^2}
\ {\rm Im} A_1 (Q^2, \nu')
\label{eqcj}
\end{equation}
Here 
$A_1$ is the first form-factor in the spin dependent part of 
the 
forward Compton amplitude
which is related to the $g_1$ spin structure function 
via $g_1 = \pi {M \over \nu} A_1$
and
$\beta_1 (Q^2)$
denotes a possible subtraction constant 
(``subtraction at infinity'')
from the circle when we close the contour in the complex plane
\cite{zakopane}.
($M$ denotes the proton mass.)

The value of $g_A^{(0)}$ extracted from polarized deep inelastic
scattering is obtained as follows.
One 
applies the operator product expansion
and
finds that the first moment of  the structure function $g_1$
is related
to the scale-invariant axial charges of the target nucleon:
\begin{eqnarray}
\int_0^1 dx \ g_1^p (x,Q^2) 
&=&
\Biggl( {1 \over 12} g_A^{(3)} + {1 \over 36} g_A^{(8)} \Biggr)
\Bigl\{1 + \sum_{\ell\geq 1} c_{{\rm NS} \ell\,}
\alpha_s^{\ell}(Q)\Bigr\} 
\nonumber \\
& & + {1 \over 9} g_A^{(0)}|_{\rm inv}
\Bigl\{1 + \sum_{\ell\geq 1} c_{{\rm S} \ell\,}
\alpha_s^{\ell}(Q)\Bigr\}
\ + \ {\cal O}({1 \over Q^2}) \ - \ \beta_1 (Q^2) {Q^2 \over 4 M^2}
\label{eqcch}
\end{eqnarray}
Here $g_A^{(3)}$, $g_A^{(8)}$ and $g_A^{(0)}|_{\rm inv}$ 
are the isovector, SU(3) octet and scale-invariant flavour-singlet 
axial charges respectively.
The Wilson coefficients $c_{{\rm NS} \ell}$ and $c_{{\rm S} \ell}$
are calculable in $\ell$-loop perturbative QCD
\cite{larin}.
One then assumes no twist-two subtraction constant
($\beta_1 (Q^2) = O(1/Q^4)$)
so that the axial charge contributions saturate the first moment
at leading twist.

The first moment of $g_1$ is constrained by low energy weak
interactions.
For proton states $|p,s\rangle$ with momentum $p_\mu$ and spin $s_\mu$
\begin{eqnarray}
2 M s_{\mu} \ g_A^{(3)} &=&
\langle p,s |
\left(\bar{u}\gamma_\mu\gamma_5u - \bar{d}\gamma_\mu\gamma_5d \right)
| p,s \rangle   \nonumber \\
2 M s_{\mu} \ g_A^{(8)} &=&
\langle p,s |
\left(\bar{u}\gamma_\mu\gamma_5u + \bar{d}\gamma_\mu\gamma_5d
                   - 2 \bar{s}\gamma_\mu\gamma_5s\right)
| p,s \rangle
\label{eqcci}
\end{eqnarray}
Here $g_A^{\scriptscriptstyle (3)} = 1.267 \pm 0.004$
is the isovector axial charge measured in neutron beta-decay;
$g_A^{\scriptscriptstyle (8)} = 0.58 \pm 0.03$
is the octet charge measured independently in
hyperon beta decays (using SU(3)) \cite{su3}.

The scale-invariant flavour-singlet axial charge $g_A^{(0)}|_{\rm inv}$
is defined by
\begin{equation}
2M s_\mu g_A^{(0)}|_{\rm inv} =
\langle p, s|
\ E(\alpha_s) J_{\mu5} \ |p, s \rangle
\label{eqccj}
\end{equation}
where
$
J_{\mu5} = \left(\bar{u}\gamma_\mu\gamma_5u
                  + \bar{d}\gamma_\mu\gamma_5d
                  + \bar{s}\gamma_\mu\gamma_5s\right)
$
is the
gauge-invariantly renormalized singlet axial-vector operator
and
$
E(\alpha_s) = \exp \int^{\alpha_s}_0 \! d{\tilde \alpha_s}\,
\gamma({\tilde \alpha_s})/\beta({\tilde \alpha_s})
$
is a renormalization group factor
which corrects
for the two-loop anomalous dimension
$\gamma(\alpha_s)$ 
of $J_{\mu 5}$
and which goes to one in the limit
$Q^2 \rightarrow \infty$.

In the isovector channel 
The Bjorken sum rule \cite{bj} for the isovector part of
$g_1$, 
$
\int_0^1 dx ( g_1^p - g_1^n )$,
has been confirmed in polarized deep inelastic scattering
experiments at the level of 10\%.
Substituting the values of $g_A^{(3)}$ and $g_A^{(8)}$
from beta-decays
(and assuming no subtraction constant correction)
in the first moment equation
(\ref{eqcch})
polarized deep inelastic data 
implies
$
g^{(0)}_A |_{\rm pDIS} = 0.15 - 0.35
$
for the flavour singlet moment.
The small $x$ extrapolation of $g_1$ data is the largest 
source of experimental error on measurements of the 
nucleon's axial charges from deep inelastic scattering.

QCD theoretical analysis leads to the formula
\cite{etar,ccm,topology}:
\begin{equation}
g_A^{(0)}
=
\biggl(
\sum_q \Delta q - 3 {\alpha_s \over 2 \pi} \Delta g \biggr)_{\rm partons}
+ {\cal C}_{\infty}
\label{eqa3}
\end{equation}
where
$g_A^{(0)} = g_A^{(0)}|_{\rm inv} / E ( \alpha_s )$
Here $\Delta g_{\rm partons}$ is the amount of spin carried 
by polarized
gluon partons in the polarized proton and
$\Delta q_{\rm partons}$ measures the spin carried by quarks 
and
antiquarks
carrying ``soft'' transverse momentum $k_t^2 \sim m^2, P^2$
where $m$ is the light quark mass and $P$ is a typical gluon
virtuality;
${\cal C}_{\infty}$ denotes a non-perturbative gluon topological 
contribution which has support only at Bjorken $x$ equal to zero
(so that it cannot be measured in polarized deep inelastic scattering)
\cite{topology}.
Since $\Delta g \sim 1/\alpha_s$ under QCD evolution, the
polarized gluon term $[-{\alpha_s \over 2 \pi} \Delta g]$
in Eq.(\ref{eqa3})
scales as $Q^2 \rightarrow \infty$ \cite{etar}.
The polarized gluon contribution corresponds to two-quark-jet
events carrying large transverse momentum $k_t \sim Q$ in 
the final state from photon-gluon fusion \cite{ccm}.

The topological term ${\cal C}_{\infty}$ may be identified with 
a leading twist 
``subtraction at infinity'' in the original dispersion relation
(\ref{eqcj}), 
whence
$g_A^{(0)}|_{\rm pDIS}$ is identified with $g_A^{(0)}-C_{\infty}$
\cite{zakopane}.
It probes the role of gluon topology in dynamical axial U(1)
symmetry breaking in the transition from current to constituent
quarks in low energy QCD.
The deep inelastic measurement of $g_A^{(0)}$, Eq.(\ref{eqa1}),
is not necessarily inconsistent with the constituent quark model
prediction 0.6
{\it if} a substantial fraction of the spin of the constituent quark
is associated with gluon topology in the transition from constituent
to current quarks  (measured in polarized deep inelastic scattering).
\footnote{
The electroweak version of this physics can lead to 
the formation of a ``topological condensate'' in 
the early Universe in parallel with the generation of
the baryon number asymmetry \cite{bass04}.}

Understanding the transverse momentum dependence of 
the contributions
to (\ref{eqa3}) 
is essential to ensure that the theory and experimental 
acceptance are correctly matched when extracting information from 
semi-inclusive measurements 
about 
the individual
valence, sea and gluonic contributions 
\cite{bass02}.
Recent semi-inclusive measurements \cite{hermes}
using a forward detector and limited acceptance at 
large transverse momentum ($k_t \sim Q$)
exhibit no evidence for the large negative polarized
strangeness polarization extracted from inclusive data 
and may, perhaps, be more comparable with $\Delta q_{\rm partons}$ 
than the inclusive measurement 
(\ref{eqa2}), which has the polarized gluon contribution included.
Further semi-inclusive measurements with increased 
luminosity and a 4$\pi$ detector would be valuable.

A direct measurement of the strange quark axial charge could
be made using neutrino proton elastic scattering.

\section{$\nu p$ elastic scattering}

Neutrino proton elastic scattering measures the proton's weak axial
charge $\gaz$ through elastic Z$^0$ exchange.
Because of anomaly cancellation in the Standard Model
the weak neutral current couples to the combination $u-d+c-s+t-b$,
{\it viz.}
\begin{equation}
J_{\mu5}^Z\
=\ \smallfrac{1}{2} \biggl\{\,\sum_{q=u,c,t} - \sum_{q=d,s,b}\,\biggr\}\:
        \bar{q}\gamma_\mu\gamma_5q
\label{eqdj}
\end{equation}
It measures the combination
\begin{equation}
2\gaz = \bigl( \Delta u - \Delta d - \Delta s \bigr)
       + \bigl( \Delta c - \Delta b + \Delta t \bigr)
\label{eqdk}
\end{equation}
where
\begin{equation}
2M s_{\mu} \Delta q =
\langle p,s |
\biggl( {\overline q} \gamma_{\mu} \gamma_5 q \biggr)
| p,s \rangle
\end{equation}
Heavy quark renormalization group arguments
can be used
to calculate the heavy $t$, $b$ and $c$ quark contributions to $\gaz$.
Working to NLO and expressing the result in terms of just 
renormalization scale invariant quantities one finds the result
\cite{bcsta}
\begin{equation}
2\gaz = \bigl(\Delta u - \Delta d - \Delta s\bigr)_{\rm inv}
           +\hsp{0.2} {\cal H}\hsp{0.1}\bigl(
               \Delta u + \Delta d + \Delta s\bigr)_{\rm inv}
    + \ O(m_{t,b,c}^{-1})
\label{eqdka}
\end{equation}
where ${\cal H}$ is a polynomial in the running couplings
$\run{h}$,
\begin{equation}
{\cal H}
 =  \smallfrac{6}{23\pi}\bigl(\run{b}-\run{t}\bigr)
             \Bigl\{1 + \smallfrac{125663}{82800\pi}\run{b}
                      + \smallfrac{6167}{3312\pi}\run{t}
                      - \smallfrac{22}{75\pi}\run{c}  \Bigr\}
- \smallfrac{6}{27\pi} \run{c}
                      - \smallfrac{181}{648 \pi^2}\run{c}^2
                      + O\bigl(\run{t,b,c}^3\bigr)
\label{eqdl}
\end{equation}
Here $(\Delta q)_{\rm inv}$ denotes the scale-invariant version of
$\Delta q$ and $\run{h}$
denotes Witten's renormalization group invariant running couplings
for heavy quark physics.
Taking $\widetilde{\alpha}_t = 0.1$, $\widetilde{\alpha}_b = 0.2$
and $\widetilde{\alpha}_c = 0.35$ in (\ref{eqdl}), one finds a small
heavy-quark correction factor ${\cal H}= -0.02$, with LO terms dominant.

Modulo the small heavy-quark corrections quoted above, a precision
measurement of $g_A^{(Z)}$,
together with
$g_A^{(3)}$ 
would
provide a rigorous weak interaction determination of 
$(\Delta s)_{\rm inv}$.
The axial charge measured in $\nu p$ elastic scattering 
is independent of any assumptions about the presence or
absence of a ``subtraction at infinity'' correction to the 
first moment of $g_1$ and the $x \sim 0$ behaviour of $g_1$.

\end{document}